\documentclass[aps,pre,twocolumn,groupedaddress]{revtex4-1}
\usepackage{amssymb}
\usepackage{amsmath}
\usepackage{graphicx}% Include figure files
\usepackage{dcolumn}% Align table columns on decimal point
\usepackage{bm}% bold math
\usepackage[english]{babel}
\usepackage{color}
\usepackage{enumitem}
\usepackage{soul}
\usepackage{epstopdf}

\newcommand{\T}{k_\mathrm{B}T}

\graphicspath{{./figures/}}  %% PATH FOR FIGURES: keep the double "{{" and "}}"

\begin{document}

\title{Thermal Equilibrium in $D$-dimensions:\\
From Fluids and Polymers to Kinetic Wealth Exchange Models}

\author{Marco Patriarca, Els Heinsalu}
%\email[]{}
%\homepage[]{Your web page}
%\thanks{}
%\altaffiliation{}
\affiliation{National Institute of Chemical Physics and Biophysics, R\"avala 10, 10143 Tallinn, Estonia\\
Email: {\tt marco.patriarca@kbfi.ee, els.heinsalu@kbfi.ee}}

\author{Amrita Singh, Anirban Chakraborti}
%\email[]{}
%\homepage[]{Your web page}
%\thanks{}
%\altaffiliation{}
\affiliation{School of Computational \& Integrative Sciences, Jawaharlal Nehru University, New Delhi 110067, India\\
Email: {\tt anirban@jnu.ac.in, s.amrita.bhu@gmail.com}}

\vspace{0.5cm} 
\date{\today}

% -----------------------------------------------------------------------
\begin{abstract}
In this paper we discuss some examples of systems composed of $N$ units, which exchange a conserved quantity $x$ according to some given stochastic rule, from some standard kinetic model of condensed matter physics to the kinetic exchange models used for studying the wealth dynamics of social systems.
The focus is on the similarity of the equilibrium state of the various examples considered, which all relax toward a canonical Gibbs-Boltzmann equilibrium distribution for the quantity $x$, given by a $\Gamma$-distribution with shape parameter $\alpha = D/2$, which implicitly defines an effective dimension $D$ of the system.
We study various systems exploring (continuous) values of $D$ in the interval $[1,\infty)$.
\end{abstract}
% -----------------------------------------------------------------------

%\keywords{}

\maketitle

% =============================================================================
\section{Introduction}
\label{introduction}
%
% Check:
% difference between Gibbs and Boltzmann (ensemble vs macroscopic sample?)

The ``canonical'' distribution $f_0(x)$  for the energy $x$ introduced in textbooks as the one characterizing systems at thermal equilibrium is the Gibbs-Boltzmann distribution~\cite{Landau5}, which has the simple form of an exponential, 
\begin{equation}
f_0(x) \sim  \exp( - \beta x) \, ,
\label{f0}
\end{equation}
where $\beta = 1 / k_\mathrm{B} T$ is the inverse absolute temperature expressed in energy units.
There are various ways of deriving such a distribution.

In practice, in many applications in which the system under study is characterized by $D$ (statistically) independent degrees of freedom, the canonical distribution assumes another shape, mathematically different from the basic form in Eq.~(\ref{f0}), but closely related to it, namely the $\Gamma$-distribution~\cite{Abramowitz1970a},
% -------------------------------------------------------------------------------------
\begin{equation}
  \gamma(x;\alpha,\beta) = \frac{ \beta(\beta x)^{\alpha-1} }{\Gamma(\alpha)} \exp\left( - \beta x \right) \, .
    \label{gamma}
\end{equation}
% -------------------------------------------------------------------------------------
The $\Gamma$-distribution is usually written in terms of the shape parameter $\alpha$ and the rate parameter $\beta$ and contains the $\Gamma$-function, $\Gamma(n) = \int_0^\infty dx \, x^{n-1}\exp(-x)$, for reasons of normalization.
The $\Gamma$-distribution (\ref{gamma}) is the universal counterpart of Eq.~(\ref{f0}) for systems with an arbitrary number $D$ of degrees of freedom, not only in the case of dynamical and statistical systems, e.g. the molecules of a $D$-dimensional gas or a $D$-dimensional oscillator in thermal equilibrium, but also in more general problems as in the $\chi^2$-distribution arising in the data analysis of a sample with $D$ data~\cite{Abramowitz1970a}.
In all these cases, the shape parameter $\alpha$ is related to the number of degrees of freedom $D$ through the relation
% -------------------------------------------------------------------------------------
\begin{equation}
  \alpha = \frac{D}{2} \, .
    \label{D}
\end{equation}
% -------------------------------------------------------------------------------------
In order to illustrate the ubiquity of the $\Gamma$-distribution --- and therefore the existence of an effective dimension $D$ characterizing many systems --- in the following we use different equivalent approaches to discuss numerically and analytically some examples, starting with the mentioned case of the molecular kinetic energy in a fluid in $D$-dimensions, the potential energy of $D$-dimensional oscillators or of a set of polymers composed by $D$ monomers, eventually to compare the results obtained with those obtained from the dynamics of the market economy described by kinetic exchange models \cite{chakrabarti2013econophysics,Sinha2010,Econo-Socio-book,Patriarca2004a,Patriarca2004b,Chakraborti-2011econophysics,Lallouache-2010}.

% =============================================================================
\section{Assembly of Polymers}
\label{polymer}
An assembly of weakly interacting harmonic polymers is a simple, yet exactly solvable model, characterized by an effective dimension $D$.
The interaction between polymers brings the system to thermal equilibrium but is otherwise assumed to be weak enough not to perturb appreciably the single polymer dynamics, so that each polymer undergoes independent statistical fluctuations coming from the environment.
If monomer-monomer interactions inside a polymer are approximated trough harmonic potentials, then a single polymer can be described by the small displacements of the normal modes around the equilibrium configuration of the system.
The number of independent normal modes represents here the effective dimension $D$.
In the expression of the potential energy distribution of a polymer, one can use rescaled coordinates of the $D$ harmonic degrees of freedom, 
$\mathbf{q} = \{q_i\} = \{q_1, q_2, \dots, q_D\}$, so that the energy function can be written in the normal form
$x(\mathbf{q}) = (q_1^2+\dots+q_D^2)/2 $.
It is clear that this problem is equivalent to that of a $D$-dimensional harmonic oscillator, so that the equilibrium solution discussed here below applies to both cases.
We start using the Gibbs approach, i.e.  from the Gibbs factor $\exp(-\beta x(\mathbf{q}))$, where $\beta$ is the inverse temperature, $\beta  =  1/\T$.
It is a simple exercise to obtain the distribution for the variable $x$.
First, one can move to $D$-dimensional coordinates and integrate out the $D - 1$ angular variables in the $D$-dimensional space $\mathbf{q}$ using the expression for the surface of the $D$-dimensional hyper-sphere of radius $r$ --- in this case the hyper-sphere is a ($D-1$)-dimensional ``surface'',
% -------------------------------------------------------------------------------------
\begin{equation}
  \label{SigmaD}
  \sigma_D(r)  = \frac{2\pi^{D/2}}{\Gamma(D/2)} \, r^{D-1} \, ,
\end{equation}
% -------------------------------------------------------------------------------------
to obtain the distribution of the modulus $F(q)$, where $q = \sqrt{\mathbf{q}^2} = \sqrt{q_1^2+\dots+q_D^2}$.
Eventually, one changes variable from $q$ to $x(q)$.
The resulting normalized distribution $f(x)$ is just the $\Gamma$-distribution of Eq.~(\ref{gamma}),
$f(x) = \beta \gamma(x;n,\beta)$, with shape parameter $n = D/2$.

It is worth noting that the perfect exponential distribution (\ref{f0}) is obtained only for the specific case of $D = 2$, while in general the exponential shape is qualitatively changed at small $x$: for $D > 2$ the distribution is zero for $x = 0$ and therefore has a mode at $x > 0$, while for $D < 2$ it presents a divergence for $x \to 0$, as discussed below in greater detail.

% =============================================================================
\section{The Maxwell velocity distribution}
\label{maxwell}
We now turn to the kinetic molecular energy distribution, which should be a universal feature of any system in which the constituent particles interact through an inter-particle potential which only depends on the space coordinates.
A most straightforward way to obtain the equilibrium distribution of the molecular kinetic energy of a fluid in $D$ dimensions, with a minimum set of assumptions, is the original derivation due to Maxwell.
The method was introduced to compute the distribution of the velocity modulus in a (3D) gas~\cite{Landau5}.
Assuming that the distributions of the velocity components $v_x$, $v_y$, $v_z$ along the $x$, $y$, and $z$-axis, respectively, are statistically independent of and equivalent to each other and that the 3-dimensional velocity distribution $f_3(v_x, v_y, v_z)$ is isotropic, only depending on the squared modulus $v^2$, one has that
$f_3(v_x, v_y, v_z) = f_3(v^2) \equiv f_3(v_x^2 + v_y^2 + v_z^2) \propto f_1(v_x) \times f_1(v_y) \times f_1(v_z)$, where the $f_1(v_i)$, $i = 1,2,3$, are the one-dimensional distributions.
This implies the following velocity distribution,  
% -------------------------------------------------------------------------------------
\begin{equation}
  \label{fv3}
  f_3(v_x, v_y, v_z) = \left(\frac{m}{2\pi T}\right)^{3/2} \exp\left[ - \frac{m\beta}{2}(v_x^2 + v_y^2 + v_z^2) \right] \, .
\end{equation}
% -------------------------------------------------------------------------------------
One can obtain the corresponding distribution of the kinetic energy modulus first by moving to spherical coordinates and integration of the angular variables and then making a final change of variable, $x = mv^2/2$.
Then one finds the kinetic energy distribution in 3D,
% -------------------------------------------------------------------------------------
\begin{equation}
  f_3(x) = \frac{2\beta^{3/2}}{\sqrt{\pi}} \sqrt{x} \exp( - \beta x ) \, .
    \label{K3D}
\end{equation}
% -------------------------------------------------------------------------------------
This method can be easily generalized for the $D$-dimensional case, proceeding in a similar way from the corresponding velocity distribution in $D$ dimensions,
% -------------------------------------------------------------------------------------
\begin{equation}
  \label{fvD}
  f_D(v_1 \dots v_D) = \left(\frac{m\beta}{2\pi}\right)^{D/2} \exp\left( - \sum_{i=1}^{D} \frac{\beta m v_i^2}{2} \right) \, ,
\end{equation}
% -------------------------------------------------------------------------------------
where $v_i$ is the velocity component along the $i$th dimension.
Introducing the velocity modulus, defined by $v^2 = \sum_{i=1}^{D} v_i^2$,
and integrating the distribution over the $D-1$ angular variables, with the help of the hyper-sphere surface $\sigma_D(r)$ given in Eq.(\ref{SigmaD}), one obtains the velocity modulus distribution,
% -------------------------------------------------------------------------------------
\begin{equation}
  f(v) = \frac{2}{\Gamma(D/2)} \left(\frac{m\beta}{2}\right)^{D/2} \, v^{D-1} \exp\left( - \frac{\beta m v^2}{2} \right) \, .
\end{equation}
% -------------------------------------------------------------------------------------
Introducing the kinetic energy $x = m v^2/2$, one  obtains again the $\Gamma$-distribution (\ref{gamma}) for $\alpha = D/2$.

% ==============================================================================================================================
\section{A variational approach}
\label{sec:variation}
As an alternative approach to the equilibrium distribution, it is worth to recall an equivalent and powerful method due to Boltzmann, based on the functional variation of the system entropy \cite{chakraborti2009}.
As above, the representative system is assumed to have $D$ degrees of freedom, $q_1,\dots,q_D$, and a homogeneous quadratic Hamiltonian $X$,
% -------------------------------------------------------------------------------------
\begin{equation}
  \label{X}
  X(q_1, \dots, q_D) \equiv X(q^2)  = \frac{1}{2} (q_1^2 + \dots + q_D^2) = \frac{1}{2} \, q^2 \, ,
\end{equation}
% -------------------------------------------------------------------------------------
where $q = (q_1^2 + \dots + q_D^2)^{1/\,2}$ is the distance from the origin in the $D$-dimensional $q$-space.
The $D$ coordinates $q_i$ can represent e.g. suitably rescaled values of the velocities [so that Eq.~(\ref{X}) provides the corresponding kinetic energy function] or the normal coordinates of a harmonic network [so that $X$ represents the total potential energy], as discussed in the previous sections.
The expression of the Boltzmann entropy of a system described by $D$ continuous variables $q_1, \dots, q_D$, is
% -------------------------------------------------------------------------------------
\begin{eqnarray}
  \label{SD}
  && S_D[q_1\dots q_D] \nonumber
  \\
  && \! = -\!\! \int \!\! dq_1\dots\int \!\! dq_D \, f_D(q_1 \dots q_D) \ln[f_D(q_1 \dots q_D)] .~~
\end{eqnarray}
% -------------------------------------------------------------------------------------
The system is subjected to the constraints on the conservations of the total number of units (meaning a normalization for a probability distribution function) and of the total wealth (implying a constant average energy $\bar{x}$),
% -------------------------------------------------------------------------------------
\begin{eqnarray}
  \label{cc}
  &&  \int dq_1\dots\int dq_D \, f_D(q_1 \dots q_D) = 1 \, ,
  \\
  &&  \int dq_1\dots\int dq_D \, f_D(q_1 \dots q_D) X(q_1, \dots, q_D) = \bar{x} \, ,
\end{eqnarray}
% -------------------------------------------------------------------------------------
Using the method of Lagrange multipliers, one should make a variation with respect to the distribution $f_D(\dots)$ of the functional
% -------------------------------------------------------------------------------------
\begin{eqnarray}
  && S_\mathrm{eff}[f_D]
  \nonumber
  \\
  && = \int  dq_1\dots\int dq_D  f_D(q_1,\dots,q_D) \{ \ln[f_D(q_1,\dots,q_D)]
  \nonumber
  \\
  && ~~~~~~~~~~~~~~~~ + \mu + \beta X(q^2) \} ,
  \label{Sn}
\end{eqnarray}
% -------------------------------------------------------------------------------------
where $\mu$ and $\beta$ are the Lagrange multipliers, but exploiting the invariance of the Hamiltonian, which depends only on the modulus $q$, it is convenient first to change from Cartesian to polar coordinates and integrate the $(D-1)$ coordinates spanning the solid angle.
Using again the expression (\ref{SigmaD}) for the surface of the hyper-sphere, one obtains
% -------------------------------------------------------------------------------------
\begin{eqnarray}
  && S_\mathrm{eff}[f_1]
  \nonumber
  \\
  &&= \int_{0}^{+\infty} dq \, f_1(q)
         \left[ \ln\left( \frac{f_1(q)}{\sigma_D^1 \, \, q^{D-1}} \right) + \mu + \beta X(q) \right] \,
  \label{func4}
\end{eqnarray}
% -------------------------------------------------------------------------------------
where $\sigma_D^1 \equiv \sigma_D(1) = (2\pi^{D/2})/\Gamma(D/2)$ is the surface of a unit-radius sphere.
Also notice that for symmetry, the probability density $f_D(q_1,\dots,q_D)$ in the $D$-dimensional space depends only on the variable $q$ and has been expressed 
in terms of the reduced probability density $f_1(q)$ in the one-dimensional $q$-space, given by
% -------------------------------------------------------------------------------------
\begin{equation}
f_1(q) = \sigma_D^1 \, \, q^{D-1} f_D(q) \, .
\end{equation}
% -------------------------------------------------------------------------------------
Finally, transforming from $q$ to the energy variable $x = q^2/\,2$, one obtains the probability distribution function
% -------------------------------------------------------------------------------------
\begin{equation}
f(x)
= \frac{dq(x)}{dx} \left.f_1(q)\right|_{q = q(x)}
= \frac{\left.f_1(q)\right|_{q = q(x)}}{\sqrt{2x}} \, ,
\end{equation}
% -------------------------------------------------------------------------------------
where $q(x) = \sqrt{2x}$ from Eq.~(\ref{X}).
In terms of the new variable $x$ and the new distribution $f(x)$ in the (1D) $x$-space, from Eq.~(\ref{func4}), one obtains the functional
% -------------------------------------------------------------------------------------
\begin{eqnarray} 
  S_\mathrm{eff}[f]
  =  \!\!\int_{0}^{+\infty} \!\!\!\!\! dx \, f(x)
     \left[
       \ln\!\left(\! \frac{f(x)}{\sigma_D^1 \, x^{D/\,2-1}} \!\right)
        \!+\! \mu
        \!+\! \beta x
      \right].~~
  \label{func6b}
\end{eqnarray}
% -------------------------------------------------------------------------------------
Varying this functional with respect to $f(x)$, $\delta S_\mathrm{eff}[f]/\delta f(x) = 0$, leads to the equilibrium $\Gamma$-distribution in Eq.(\ref{gamma}) with shape parameter $\alpha = D/\,2$.
As it is clear from the discussion presented above, the only conditions for obtaining a canonical distribution are the conservation of the number of constituent units and the global conservation of a quantity $x$ exchanged between the units.
Therefore we have to expect to find that the canonical distribution characterizes much more general types of systems.
In the following section we consider such an example, originally developed for modeling economic systems.

% ==============================================================================================================================
\section{Kinetic Exchange Models}
\label{KEMs}
Kinetic exchange models describe systems of $N$ interacting units which exchange a conserved quantity $x$~\cite{Patriarca2010b}.
There are many possible interpretation for these models~\cite{Patriarca2013a}.
They were originally introduced as models of economy in which $x$ represents money or wealth.
Recently, they were also used in the study of opinion dynamics as well as in condensed matter physics~\cite{Patriarca2013a}.

Kinetic exchange models are becoming more and more popular also as prototypical statistical mechanical models of (energy) exchange.
In fact, their equilibrium state is described by a Boltzmann distribution in $D$ dimensions, i.e. by the $\Gamma$-distribution $\gamma(x;\alpha,\beta)$, with the additional peculiar feature that by tuning some parameters of the model one can vary the value of the effective dimension $D$ in a continuous way in the interval $D = (1, \infty)$.
The actual shape of the equilibrium distribution in kinetic exchange models is still an active subject of investigation.

In the basic versions of kinetic exchange models, $N$ agents exchange a quantity $x$ which represents the wealth.
The state of the system is characterized by the set of variables $\{x_i\},~i = 1, 2, \dots, N$, the wealths of the $N$ agents.
In the basic version of the models, in which the evolution of the system proceeds through pair-wise interactions, as well as in the more general ones, in which an interaction can involve all $N$ units, the total wealth is conserved during each interaction,
\begin{equation}
X = x_1 + x_2 + \cdots + x_{N-1} + x_N = \mathrm{const} .
\label{XP}
\end{equation} 
The time evolution is carried out through a prescription --- the update rule --- which has to be assigned.
For convenience, in this paper we consider the basic version of the ``immediate-exchange models'', introduced in Ref.~\cite{Heinsalu2014c,Heinsalu2015a}. 
In the homogeneous version of this model, at every time step $t$, two agents $i$ and $j$ are extracted randomly and, according to the following rule, a random redistribution of the money takes place,
\begin{eqnarray}
  &  x_i \to x_i' &= \lambda x_i + (1 - \lambda) (- \epsilon_1 x_i  +  \epsilon_2 x_j) , \nonumber
  \\
  &  x_j \to x_j' &= \lambda x_j + (1 - \lambda) (  \epsilon_1 x_i  -  \epsilon_2 x_j) .
  \label{basic0}
\end{eqnarray}
Here $x_i'$ and $x_j'$ are the wealths of the units after an interaction and $\lambda$, the saving parameter, represents the fraction of wealth saved at each interaction, while $(1 - \lambda)$ is the complementary fraction reshuffled randomly between the two units $i$ and $j$.
A feature of immediate-exchange models is the dependence of the dynamics on two independent random numbers $\epsilon_1$ and $\epsilon_2$, describing the independent random fluctuations affecting the behavior and choices of the two interacting units.
Assuming that $\epsilon_1$ and $\epsilon_2$ are uniformly distributed in the interval $\epsilon = (0,1)$, after a large number of iterations the system relaxes toward an equilibrium state characterized by a wealth distribution $f(x)$ which numerical experiments show to be well fitted by the $\Gamma$-function (\ref{gamma}) with scale and shape parameters
\begin{eqnarray}
  && \beta  = D/\langle x \rangle \, ,
  \\
  && D = 2\alpha = \frac{1 + 2\lambda}{1 - \lambda} \, .
\label{exp}
\end{eqnarray}
where $\langle x \rangle$ is the average wealth of the system.
Therefore the effective dimension $D$ of the system has its minimum value at $D = 1$, corresponding to a zero saving parameter, $\lambda = 0$.
As $\lambda$ grows from $\lambda = 0$ toward $\lambda = 1$, $D$ also grows, thus exploring the whole interval of dimensions $D \ge 1$, eventually diverging for $\lambda \to 1$.
Until recently, the quality of the numerical fitting of the results of numerical simulations was the only argumentation for claiming that kinetic exchange models relax toward canonical distributions.
However, recently some exact solutions are being found.
For instance, Katriel has shown, at least in a particular case, that the immediate-exchange model discussed above has indeed  an equilibrium $\Gamma$-distribution~\cite{Katriel2015b}.

% =============================================================================
\section{Kinetic theory in $D$-dimensions}
\label{sec:kinetic}
While the Boltzmann approach shows that conservation of the total wealth $x$ and of the total number of agents $N$ are sufficient conditions for the equilibrium state to be described by canonical distributions, it is instructive to check how the deep analogy between the dynamics of kinetic exchange models,  in which the constituent units exchange a quantity $x$, and kinetic gas models, in which particles exchange energy at each collision, is valid also at the microscopic level of single pair-wise interactions.
The analogy between trades and molecular collisions was clearly noticed by Mandelbrot~\cite{Mandelbrot1960a}, but it can been best illustrated by showing the equivalence between the microscopic dynamics of the standard kinetic theory of gases and the update rules of kinetic exchange models.

In a 1D gas, in the absence of external noise, particles undergo head-on collisions and simply exchange their kinetic energies at each collision.
In this way, the energy distribution does not evolve in time.

To have energy redistributed among particles during the collisions, one has to go at least to a 2-dimensional space or a general $D$-dimensional space with $D \ge 2$.
Then, there is in general no head on collision unless the two particles are traveling exactly along the same line in opposite verses.
On average, only a fraction of the total kinetic energy of a molecule will be lost or gained during a collision.

We then consider a collision between two particles in an $D$-dimensional space,
with initial velocities represented by the vectors ${\bf v}_{(1)} = (v_{(1)1}, \dots, v_{(1)D})$ and ${\bf v}_{(2)} = (v_{(2)1}, \dots, v_{(2)D})$.
For the sake of simplicity, the masses of the two particles are assumed to be equal to each other and is set equal to 1, so that momentum conservation implies that
\begin{eqnarray}  \label{v1}
  {\bf v}_{(1)}' &=& {\bf v}_{(1)} + \Delta {\bf v} \, ,
  \nonumber \\
  {\bf v}_{(2)}' &=& {\bf v}_{(2)} - \Delta {\bf v} \, ,
\end{eqnarray}
where ${\bf v}_{(1)}'$ and ${\bf v}_{(2)}'$ are the velocities after the collisions and $\Delta {\bf v}$ is the momentum transferred.
Conservation of energy implies that ${\bf v}_{(1)}'^{\,2} + {\bf v}_{(2)}'^{\,2} = {\bf v}_{(1)}^2 + {\bf v}_{(2)}^2$ which, by using Eq.~(\ref{v1}), leads to
\begin{eqnarray}  \label{v2}
  \Delta {\bf v}^2 + ({\bf v}_{(1)} - {\bf v}_{(2)}) \cdot \Delta {\bf v} = 0 \, .
\end{eqnarray}
Introducing the cosines $r_i$ of the angles $\alpha_i$ between the momentum transferred $\Delta {\bf v}$ and the initial velocity ${\bf v}_{(i)}$ of the $i$-th particle ($i = 1,2$),
\begin{eqnarray}  \label{cos}
  r_i = \cos\alpha_i  = \frac{{\bf v}_{(i)} \cdot \Delta {\bf v}}{v_{(i)} \, \Delta v} \, ,
\end{eqnarray}
where $v_{(i)} = |{\bf v}_{(i)}|$ and  $\Delta v =  |\Delta {\bf v}|$, and using Eq.~(\ref{v2}), one obtains that the modulus of momentum transferred is
\begin{eqnarray}  \label{v3}
  \Delta v = - r_1 v_{(1)} + r_2 v_{(2)} \, .
\end{eqnarray}
From this expression one can now compute explicitly the differences in particle energies $x_i$ due to a collision,
that are the quantities $x_i' - x_i \equiv ({\bf v}_{(i)}'^{\,2} - {\bf v}_{(i)}^2)/\,2$.
With the help of the relation (\ref{v2}) one obtains
\begin{eqnarray}  \label{x1}
  x_1' &=& x_1 + r_2^2 \, x_2 - r_1^2 \, x_1 \, ,
  \nonumber \\
  x_2' &=& x_2 - r_2^2 \, x_2 + r_1^2 \, x_1 \, .
\end{eqnarray}
The equivalence to kinetic exchange models now appears clearly.
First, the number $r_i$'s are squared cosines and therefore they are in the interval $r \in (0,1)$.
Furthermore, they define the initial directions of the two particles entering the collision, so that they can be considered as random variables if the hypothesis of molecular chaos is assumed.
In this way, they are analogous to the random coefficients $\epsilon_i(1-\lambda)$ ($i = 1,2$) appearing above in the formulation of kinetic exchange models, with the difference that the latter ones cannot assume all values in $(0,1)$ but are limited in the interval $(0,1-\lambda)$. 
However, in general the $r_i$'s are not uniformly distributed in $(0,1)$ and their most probable values $\langle r_i^2 \rangle$ drastically depend on the space dimension, which is at the base of their effective equivalence with the kinetic exchange models: the greater the dimension $D$, the smaller the $\langle r_i^2 \rangle$, since the more unlikely it becomes that the corresponding values $\langle r_i \rangle$ assume values close to 1 and the more probable that instead they assume a small value close to $\sim 1/D$.
This can be seen by computing their average -- over the incoming directions of the two particles or, equivalently, on the orientation of the initial velocity ${\bf v}_{(i)}$ of one of the two particles and of the momentum transferred $\Delta {\bf v}$, which is of the order of $1/D$.

Kinetic exchange models can be studied not only numerically but in some cases an analytical solution can be obtained.

% =============================================================================
\section{Molecular dynamics in $D$-dimensions}
\label{sec:nD}
%

%%%%%%%%%%%% FIGURE %%%%%%%%%%%%
\begin{figure}[ht]
\centering
\includegraphics[width=7cm]{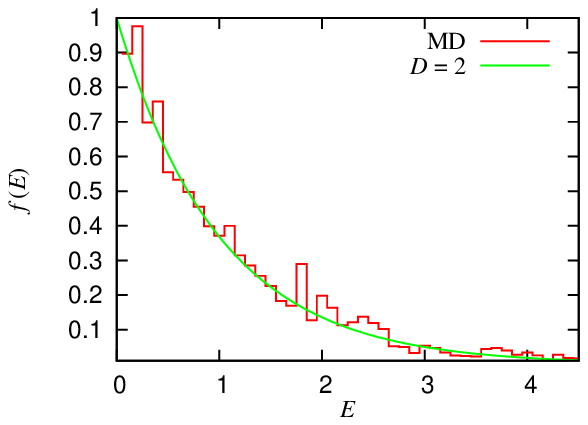},
\includegraphics[width=7cm]{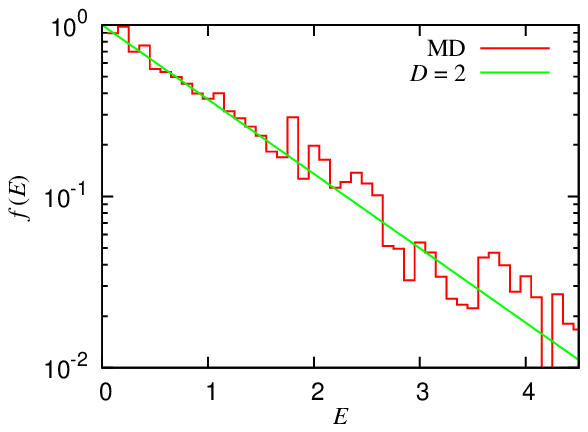}
\caption{Example of kinetic energy distribution for a gas in $D = 2$ dimensions in the linear (top) and linear-logarithmic (bottom) scale.
The ``MD'' histogram represents the numerical result of the MD simulation, while the continuous curve $D = 2$ is the theoretical equilibrium distribution, given by the $\Gamma$-distribution with shape parameter $\alpha = D/2$ and a scale parameter here given by $\theta = T = 1$.
Notice that only in the specific case of $D = 2$ dimensions the equilibrium distribution is a perfect exponential.}
\label{fig_D2}
\end{figure}
%%%%%%%%%%%%%%%%%%%%%%%%%%%%%%%%%%%%%%%%%

It can be instructive, to close our review of approaches to equilibrium in $D$ dimensions, to discussing a different but equivalent method, 
namely to use molecular dynamics (MD) simulations to study directly the relaxation to equilibrium of a $D$-dimensional gas.
For the sake of simplicity we consider the distribution of kinetic energy of the gas, since it known to relax to the Boltzmann distribution with 
the proper number $D$ of dimensions of the gas \emph{independently of the inter-particle potential}.

%%%%%%%%%%%% FIGURE %%%%%%%%%%%%
\begin{figure}[ht]
\centering
\includegraphics[width=7cm]{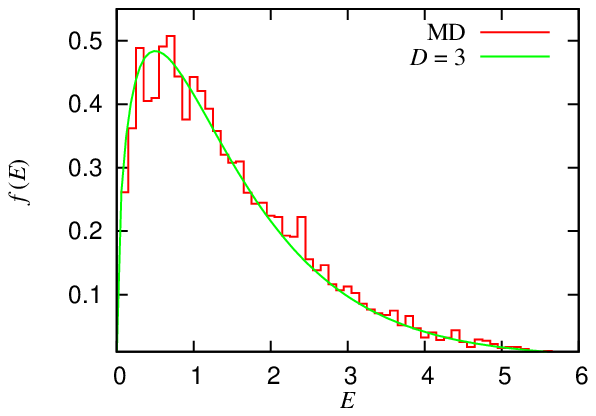}
\includegraphics[width=7cm]{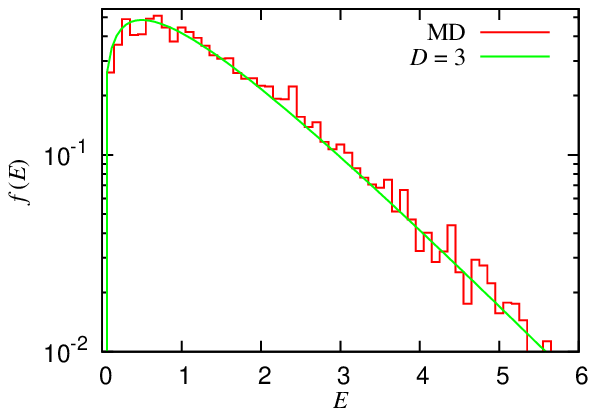}
\caption{As in the previous figure, but in $D = 3$ dimensions.}
\label{fig_D3}
\end{figure}
%%%%%%%%%%%%%%%%%%%%%%%%%%%%%%%%%%%%%%%%%
 
%%%%%%%%%%%% FIGURE %%%%%%%%%%%%
\begin{figure}[ht]
\centering
\includegraphics[width=7cm]{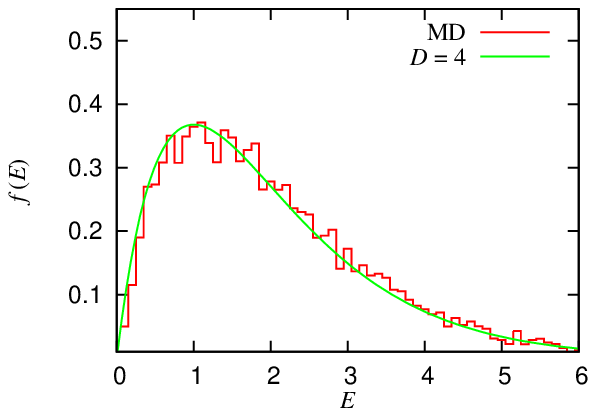}
\includegraphics[width=7cm]{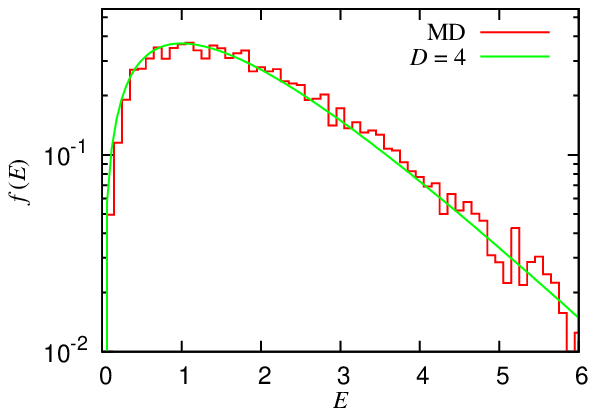}
\caption{As in the previous figure, but in $D = 4$ dimensions.}
\label{fig_D4}
\end{figure}
%%%%%%%%%%%%%%%%%%%%%%%%%%%%%%%%%%%%%%%%%

We have performed numerical simulation of a gas in a space with dimension $D$, for various values of $D$, using the leapfrog algorithm~\cite{Frenkel1996a}.
Reflecting boundary conditions for the cubic simulation (hyper-)box were used
and a repulsive Lennard-Jones pair-wise interaction potential $U(r)$, defined by  
\begin{eqnarray}  \label{U}
  U(r) && = \epsilon \left[ \left( R / r \right)^6 - 1 \right]^2 \, \mathrm{for} ~~~r < R \, , \nonumber \\ 
  \,   && = 0 \, ~~~~~~~~~~~~~~~ \mathrm{for} ~~~r \ge R \, .
\end{eqnarray}
was assumed, where $r$ is the inter-particle distance in $D$ dimensions.
This formula describes a purely repulsive potential, decreasing monotonously as the inter-particle distance $r$ increases from small distances up to a distance $R$, where the potential becomes equal to zero and then remains zero for $r > R$.
Details about the code will be presented elsewhere.

%%%%%%%%%%%% FIGURE %%%%%%%%%%%%
\begin{figure}[ht]
\centering
\includegraphics[width=7cm]{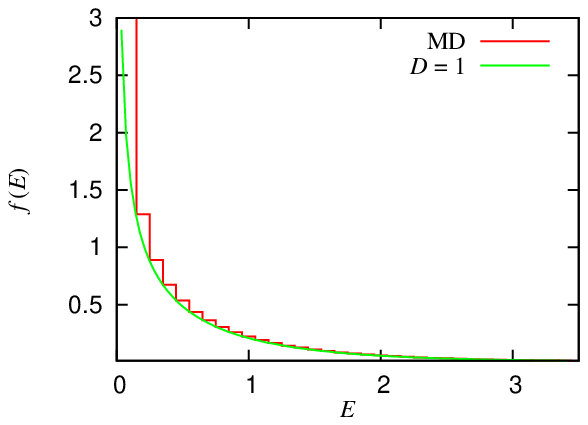}
\includegraphics[width=7cm]{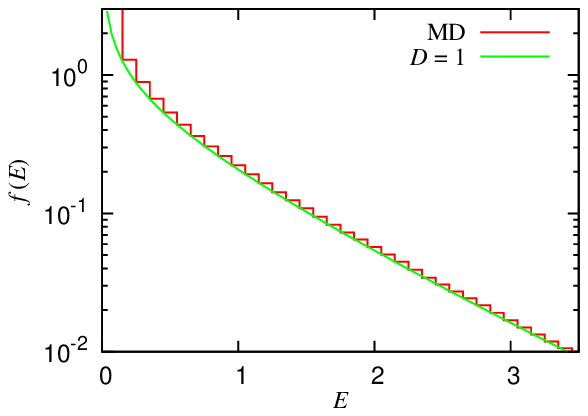}
\caption{Same as in the previous figure, but in $D = 1$ dimension.}
\label{fig_D1}
\end{figure}
%%%%%%%%%%%%%%%%%%%%%%%%%%%%%%%%%%%%%%%%%

For clarity we start by considering the case $D = 2$ dimensions.
In this case the shape parameter of the $\Gamma$-function is $\alpha = D/2 = 1$ and the equilibrium energy distribution is a perfect exponential function, $f(x) = \langle x \rangle^{-1} \, \exp( - x / \langle x \rangle)$.
We have simulated a small system consisting of $N = 20$ particles moving in a square box with a rescaled size $L = 10$.
The numerical simulation was made for a simulation time $t_\mathrm{tot} = 10^4$, using an integration time step $\delta t = 10^{-4}$ and finally computing the histogram of the kinetic energy distribution.
The histogram was then averaged over other $10^5$ snapshots equidistant in time.
The results for the kinetic energy distribution in $D = 2$ are shown in Fig.~\ref{fig_D2} both in linear and linear-logarithmic scale, the latter showing clearly the presence of the Boltzmann exponential tail.

As the number of dimensions grows, the number of particles (for constant simulation time) or the simulation time (for the same number of particles) necessary to get significant statistics grows faster than $D$.
Considering this fact, we have suitably changed some parameters of when performing numerical simulations of a gas in a cubic box in $D = 3$ dimensions and a hyper-box in $D = 4$ dimensions.
The form and parameters of the pair-wise potential is the same, apart from the corresponding generalization to higher dimensions.
Results are presented in Fig.~\ref{fig_D3} and \ref{fig_D4}.

It is instructive to consider also the one-dimensional case, $D = 1$.
Notice that in $D = 1$, using Newtonian dynamics in the MD simulation makes the velocity distribution remain unchanged in time (in a homogeneous gas), since in each collision the two colliding particles simply exchange their momenta.
Therefore, the case $D = 2$ represents the one with the minimum possible number of dimensions in which one can study the relaxation to thermal equilibrium by using energy-conserving dynamics.
Instead, to achieve thermalization in the 1-$D$ case, we have used a Langevin thermostat with rescaled temperature $T = 1$ and damping coefficient $\eta = 0.5$, in order to induce a redistribution of the kinetic energies and the thermalization of the system.

Comparison of the figures corresponding to different dimensions shows the apparent features of the corresponding equilibrium distributions depending on $D$, namely the mode and the limit $f(x \to 0) \to 0$ for dimensions $D > 2$, the pure exponential shape for the case $D = 2$, and, for $D < 2$, in the results obtained from the Langevin dynamics in $D = 1$, Fig.~\ref{fig_D1}, the divergence of the probability distribution function in the origin.

% =============================================================================
\section{Conclusion}
\label{conclusions}

We have discussed a set of examples corresponding to different types of systems, all described by a canonical equilibrium distribution at equilibrium, which are characterized by an effective dimensions $D$ of the system.
In all these cases, the probability distribution function is a $\Gamma$-distribution with shape parameter $\alpha = D/2$.
We tried to illustrate the ubiquity of the $\Gamma$-distribution through a comparison among the various systems, looking at the common points and similarities while using different, complementary points of view, within an analytical and/or a numerical approach.
All together, the examples considered show how the Boltzmann variational principle applies equally well in various systems from different fields  such as condensed matter and social sciences, characterized by a fixed number of units $N$ exchanging a conserved quantity $x$.

% --------------------------------------------------------------------------------------------------
\begin{acknowledgments}
M.P., E.H., and A.C. acknowledge support from the Institutional Research Funding IUT (IUT39-1) of the Estonian Ministry of Education and Research.
A.C. acknowledges financial support from grant number BT/BI/03/004/2003(C) of Government of India, Ministry of Science and Technology, Department of Biotechnology, Bioinformatics Division.
A.S. is grateful to Council of Scientific and Industrial Research (CSIR), New Delhi, India for the financial support.
\end{acknowledgments}
% ---------------------------------------------------------------------------------------------------

% --------------------------------------------------------------
\bibliographystyle{plain}
\bibliography{KEM-nDimensions}
%\input{xxx.bbl}
% --------------------------------------------------------------

\end{document}